\documentclass[final,3p,times,twocolumn,authoryear]{elsarticle}
\usepackage{amssymb}
\usepackage{amsthm}

\newcommand\arcsec{\hbox{$^{\prime\prime}$}}
\newcommand\farcs{\hbox{$.\!\!^{\prime\prime}$}}

\def\astrobj#1{#1}

\journal{New Astronomy}

\begin{document}

\begin{frontmatter}

%% Title, authors and addresses

%% use the tnoteref command within \title for footnotes;
%% use the tnotetext command for theassociated footnote;
%% use the fnref command within \author or \address for footnotes;
%% use the fntext command for theassociated footnote;
%% use the corref command within \author for corresponding author footnotes;
%% use the cortext command for theassociated footnote;
%% use the ead command for the email address,
%% and the form \ead[url] for the home page:
%% \title{Title\tnoteref{label1}}
%% \tnotetext[label1]{}
%% \author{Name\corref{cor1}\fnref{label2}}
%% \ead{email address}
%% \ead[url]{home page}
%% \fntext[label2]{}
%% \cortext[cor1]{}
%% \address{Address\fnref{label3}}
%% \fntext[label3]{}

\title{The magnetic variability of the $\beta$~Cep star $\xi^{1}$\,CMa}

%% use optional labels to link authors explicitly to addresses:
%% \author[label1,label2]{}
%% \address[label1]{}
%% \address[label2]{}

\author[aip]{S.~P.~J\"arvinen\corref{cor1}}
\ead{sjarvinen@aip.de}
\cortext[cor1]{Corresponding author}

\author[aip]{S.~Hubrig}
\author[eso]{M.~Sch\"oller}
\author[aip]{I.~Ilyin}

\address[aip]{Leibniz-Institut f\"ur Astrophysik Potsdam (AIP), An der Sternwarte~16, 14482~Potsdam, Germany}

\address[eso]{European Southern Observatory, Karl-Schwarzschild-Str.~2, 85748~Garching, Germany}

%\author{}
%\address{}

\begin{abstract}

  $\xi^{1}$\,CMa is a known magnetic star showing
  rotationally modulated magnetic variability with a period of 2.17937\,d.
  However, recent work based on high-resolution spectropolarimetry suggests
  that the rotation period is longer than 30\,years. We compare our new
  spectropolarimetric measurements with FORS\,2 at the VLT acquired on three
  consecutive nights in 2017 to previous FORS\,1/2 measurements of the
  longitudinal magnetic field strength. The new longitudinal magnetic field
  values are in the range from 115 to 240\,G and do not support the presence
  of a long period.

\end{abstract}


\end{frontmatter}

%% main text

\section{Introduction}
\label{intro}

\astrobj{$\xi^{1}$\,CMa} is known to be a bright (m$_{\rm V}=4.3$) sharp-lined
$\beta$~Cephei variable with spectral type B0.5--B1\,IV
\citep[e.g.][]{morel2008}.
$\beta$~Cephei variables are massive (7 to 20\,solar masses), B0--B3 
main-sequence stars pulsating in several pressure and gravity modes with 
periods between 3 and 8\,hours. Most uncertainties in the stellar structure 
of these massive stars are related to the amount of core convective 
overshooting and rotational mixing. The lack of knowledge on the internal
structure of these progenitors of core-collapse supernovae has large
consequences for the whole duration and the end of the stellar life, and,
consequently, for any astrophysical results relying on it, such as the
chemical enrichment of galaxies. The presence of large-scale organised
magnetic fields in $\beta$~Cephei stars is well established
\citep[e.g.][]{Hubrig2006, Hubrig2009, Briquet2012}.
It is also generally accepted that these magnetic fields are frozen-in fields
that were acquired during stellar formation on an early pre-main sequence
stage. Pulsating stars with magnetic fields are considered as promising
targets for asteroseismic analysis
\citep[e.g.][]{shiaerts},
which requires as input the observed parameters of the magnetic field
geometry.

To determine the magnetic field geometry and its strength, the rotation
period has to be known. However, only for very few $\beta$~Cephei stars has
the behaviour of the magnetic field been studied over the rotation cycle.
Using low-resolution spectropolarimetric observations with the FOcal Reducer
low dispersion Spectrograph
\citep[FORS\,1/2;][]{messenger}
in spectropolarimetric mode at the Very Large Telescope (VLT) and the
Soviet-Finnish echelle spectrograph
\citep[SOFIN;][]{tuominen}
installed  at the 2.56\,m Nordic Optical Telescope on La Palma,
\citet{Hubrig2006, Hubrig2009}
were able to detect magnetic fields in a few $\beta$~Cephei stars. The first
detection of a magnetic field in  $\xi^{1}$\,CMa was achieved by
\citet{Hubrig2006}.

Among the targets with detected magnetic fields, a few $\beta$~Cephei
variables were selected for follow-up VLT multi-epoch magnetic field
measurements with the aim to detect their most probable periods and to put
constraints on their magnetic field geometry
\citep{Hubrig2011a, Hubrig2011b}.
The simplest models of the magnetic field geometry are usually based on the
assumption that the studied stars are oblique dipole rotators, i.e.\ their
magnetic field can be approximated by a dipole with the magnetic axis inclined
to the rotation axis. Using the dipole approximation,
\citet{Hubrig2011a}
showed that $\xi^{1}$\,CMa is a rather fast rotating star with a rotation
period of 2.17937$\pm$0.00012\,d seen close to pole-on and having a rather
strong dipole magnetic field of the order of 5.3\,kG.

A few years later,
\citet{Fossati}
presented three additional measurements of $\xi^{1}$\,CMa obtained with
FORS\,2 in 2013 and suggested a rotation period of 2.17950\,d. On the other
hand,
\citet{shultz}
used high-resolution ($R\approx67\,000$) spectropolarimetric observations
made with the Multi-Site Continuous Spectroscopy 
\citep[MuSiCoS;][]{musicos} 
between 2000 and 2002 and with the Echelle SpectroPolarimetric Device for the 
Observation of Stars
\citep[ESPaDOnS;][]{espadons, espperf} 
between 2008 and February 2017 and proposed that the longitudinal magnetic 
field is gradually decreasing since 2010. Correspondingly, the rotation 
period of $\xi^{1}$\,CMa should be longer than 30\,years. Notably, the 
determination of this long rotation period was partly based on the MuSiCoS 
observations that were carried out at an airmass never below 2.45. Only these 
observations show the presence of a longitudinal magnetic field with negative 
polarity, whereas all other measurements with both FORS\,1/2 and ESPaDOnS 
exhibit positive polarity. To verify the proper operations of MuSiCoS during 
that period,
\citet{shultz}
refer to observations of \astrobj{36\,Lyn}. However, since 36\,Lyn is an
object that crosses Pic-du-Midi close to zenith, it is rather questionable if
this star can actually support the validity of the results obtained on
$\xi^{1}$\,CMa. For example,
\citet{Petit2011},
who used NARVAL and ESPaDOnS spectropolarimetric observations for a sensitive
magnetic field search in Sirius, dismissed all observations obtained at
high airmass from their study. Therefore, it is not clear whether the
detection of a magnetic field of negative polarity using MuSiCoS at high
airmass can be considered genuine.

Since previous low-resolution FORS\,2 observations did not show any hint at a
long-term periodicity, we obtained three additional low-resolution FORS\,2
observations for this target on three consecutive nights in October 2017, to
investigate the possible presence of a much longer rotation period. In the
following, we discuss the results of our analysis of these three new
observations and compare all currently available FORS\,1/2 measurements of
$\xi^{1}$\,CMa with those obtained using high-resolution spectropolarimetry.

\section{Observations and magnetic field analysis}
\label{obs}

The FORS\,2 multi-mode instrument is equipped with polarisation analysing
optics comprising super-achromatic half-wave and quarter-wave phase retarder
plates, and a Wollaston prism with a beam divergence of 22$\arcsec$ in
standard resolution mode. We used the GRISM 600B and the narrowest available
slit width of 0$\farcs$4 to obtain a spectral resolving power of
$R\approx2000$. The observed spectral range from 3250 to 6215\,\AA{} includes
all Balmer lines, apart from H$\alpha$, and numerous helium lines. For the
observations, we used a non-standard readout mode with low gain
(200kHz,1$\times$1,low), which provides a broader dynamic range, hence 
allowing us to reach a higher signal-to-noise ratio (S/N) in the individual 
spectra.

A description of the assessment of the presence of a longitudinal magnetic
field using FORS\,1/2 spectropolarimetric observations was presented in our
previous work
\citep[e.g.][and references therein]{Hubrig2004a, Hubrig2004b}.
Rectification of the $V/I$ spectra was performed in the way described by
\citet{Hubrig2014}.
Null profiles, $N$, are calculated as pairwise differences from all available
$V$ profiles so that the real polarisation signal should cancel out. From
these, 3$\sigma$-outliers are identified and used to clip the $V$ profiles.
This removes spurious signals, which mostly come from cosmic rays, and also
reduces the noise. A full description of the updated data reduction and
analysis will be presented in a separate paper
\citep[Sch\"oller et al., in preparation, see also][]{Hubrig2014}.
The mean longitudinal magnetic field, $\left< B_{\rm z}\right>$, is measured on
the rectified and clipped spectra based on the relation following the method
suggested by
\citet{angelland}:

\begin{eqnarray}
\frac{V}{I} = -\frac{g_{\rm eff}\, e \,\lambda^2}{4\pi\,m_{\rm e}\,c^2}\,
\frac{1}{I}\,\frac{{\rm d}I}{{\rm d}\lambda} \left<B_{\rm z}\right>\, ,
\label{eqn:vi}
\end{eqnarray}

\noindent
where $V$ is the Stokes parameter that measures the circular polarization,
$I$ is the intensity in the unpolarized spectrum, $g_{\rm eff}$ is the effective
Land\'e factor, $e$ is the electron charge, $\lambda$ is the wavelength,
$m_{\rm e}$ is the electron mass, $c$ is the speed of light,
${{\rm d}I/{\rm d}\lambda}$ is the wavelength derivative of Stokes~$I$, and
$\left<B_{\rm z}\right>$ is the mean longitudinal (line-of-sight) magnetic field.

%---------------------- F1
\begin{figure}
  \centering
  \includegraphics[width=.45\textwidth]{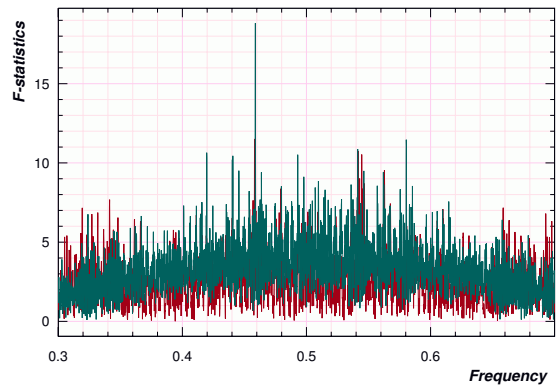}
  \caption{
    F-statistics periodogram for the longitudinal magnetic field measurements
    of $\xi^{1}$\,CMa using the entire spectrum, published in the work by
    \citet{Hubrig2011a}.
    The window function is presented in red.
           }
   \label{fig:Fstat}
\end{figure}

The longitudinal magnetic field was measured in two ways: using the entire
spectrum including all available lines or using exclusively hydrogen lines.
Furthermore, we have carried out Monte Carlo bootstrapping tests. These are
most often applied with the purpose of deriving robust estimates of standard
errors
\citep[e.g.][]{Steffen2014}.
The measurement uncertainties obtained before and after the Monte Carlo
bootstrapping tests were found to be in close agreement, indicating the
absence of reduction flaws. In Fig.~\ref{fig:Fstat} we show the periodogram
calculated from the measurements presented by
\citet{Hubrig2011a}
using the entire spectrum and utilizing the Levenberg--Marquardt method
\citep{press},
and which was not presented in this previous work.

%---------------------- T1
\begin{table*}
\centering
\caption{
  New longitudinal magnetic field measurements of $\xi^{1}$\,CMa obtained
  using FORS\,2 observations.
}
\label{tab:mg}
\begin{tabular}{cccr@{$\pm$}lr@{$\pm$}lr@{$\pm$}lc}
\hline\hline
\multicolumn{1}{c}{HJD} &
\multicolumn{1}{c}{Date} &
\multicolumn{1}{c}{S/N} &
\multicolumn{2}{c}{$\left<B_{\rm z}\right>_{\rm all}$} &
\multicolumn{2}{c}{$\left<B_{\rm z}\right>_{\rm hyd}$} &
\multicolumn{2}{c}{$\left<B_{\rm z}\right>_{\rm N}$} &
\multicolumn{1}{c}{Phase} \\
2\,450\,000+&
&
\multicolumn{1}{c}{at 4700\AA{}} &
\multicolumn{2}{c}{[G]} &
\multicolumn{2}{c}{[G]} &
\multicolumn{2}{c}{[G]} &
 \\
\hline
8029.810749 & 2017--10--03 & 1875 & 116 & 69 & 115 & 102 & $-$31 & 59 & 0.42 \\
8030.790235 & 2017--10--04 & 2610 & 181 & 57 & 240 & 77  & $-$56 & 61 & 0.87 \\
8031.740386 & 2017--10--05 & 2720 & 157 & 76 & 207 & 95  & 45    & 65 & 0.30 \\
\hline
\end{tabular}
\end{table*}

For the new observations, we present in Table~\ref{tab:mg} the heliocentric
Julian date together with the observing date, followed by the achieved
signal-to-noise ratio, and the results of the longitudinal magnetic field
measurements obtained again in two ways: using the entire spectrum or
exclusively the hydrogen lines. The measurements using the diagnostic null
spectra $N$ calculated using the entire spectrum and the rotational phases
corresponding to $P_{\rm rot}=2.17937$\,d are presented in the last two columns.

We note that low-resolution spectropolarimetric observations are most
appropriate for the detection of weak magnetic fields in $\beta$~Cep pulsating
stars, for which the impact of short-time variability on the magnetic field
measurements can be quite significant. The work of
\citet{silva}
indicated that ignoring the velocity shifts leads to an underestimation of the
magnetic field strength roughly by a factor of two. Furthermore, the
pulsational radial velocity amplitude of the order of 70~km\,s$^{-1}$ for
$\xi^{1}$\,CMa is one of the largest known among $\beta$~Cep stars
\citep[e.g.][]{saesen}.
While a full HARPS sequence of subexposures for $\xi^{1}$\,CMa requires about
30\,min, one FORS\,2 observation of the same star lasts less than 10\,min.
Owing to the strong changes in the positions and shapes of line profiles in
the spectra of pulsating stars, a method using high-resolution spectra
averaged over all subexposures usually leads to erroneous wavelength shifts
and thus to wrong values for the longitudinal magnetic field. Interestingly,
\citet{shultz}
report on the detection of apparent scatter in the measurements of the mean
longitudinal magnetic field with a standard deviation of up to 15\,G. To
explain this scatter, the authors suggest the presence of a modulation of the
field with pulsation phase or some unidentified instrumental effect, but also
consider the possibility of the impact of line profile variability.

%---------------------- F2
\begin{figure}
  \centering
  \includegraphics[width=.45\textwidth]{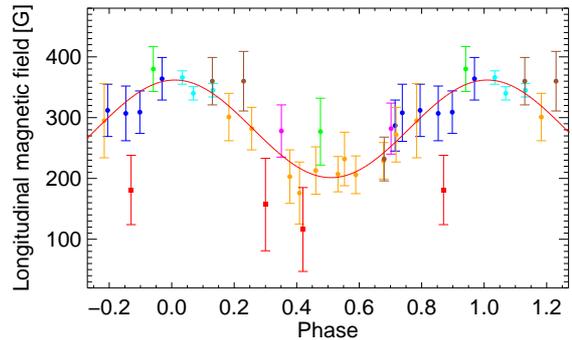}
  \caption{
    Phase folded measurements of the longitudinal magnetic field for
    $\xi^{1}$\,CMa. The filled circles represent previously published values
    from
    \citet{Hubrig2006, Hubrig2011a}
    and
    \citet{Fossati}.
    The different colors correspond to the different time bins presented in
    Fig.~\ref{fig:Btime}. The red filled squares are the new measurement
    values. The sinusoidal fit parameters and the ephemeris
    MJD$55140.73332 \pm 0.03794 + 2.17937 \pm 0.00012E$ used for the folding
    of the phase are from
    \citet{Hubrig2011a}.
           }
   \label{fig:Bphased}
\end{figure}

In Fig.~\ref{fig:Bphased} we present the phase diagram for the magnetic field
measurements of $\xi^{1}$\,CMa, using the period of 2.17937\,d, including the
three recent FORS\,2 measurements together with the results from
\citet{Hubrig2006, Hubrig2011a}
and
\citet{Fossati}.
Our new measurements acquired on three consecutive nights in 2017 indicate a
longitudinal magnetic field strength in the range from 115 to 240\,G. Two
measurements have rather large uncertainties and still fit the magnetic phase
curve well. Only the measurement obtained on 2017 October 4
($\left<B_{\rm z}\right>_{\rm all}=181\pm57$\,G,
($\left<B_{\rm z}\right>_{\rm hyd}=240\pm77$\,G) has a significance level of
about 3$\sigma$. Therefore, the obtained new observations do not provide much
evidence for the presence of the period of more than 30 years reported by
\citet{shultz}.

\begin{figure}
  \centering
  \includegraphics[width=.45\textwidth]{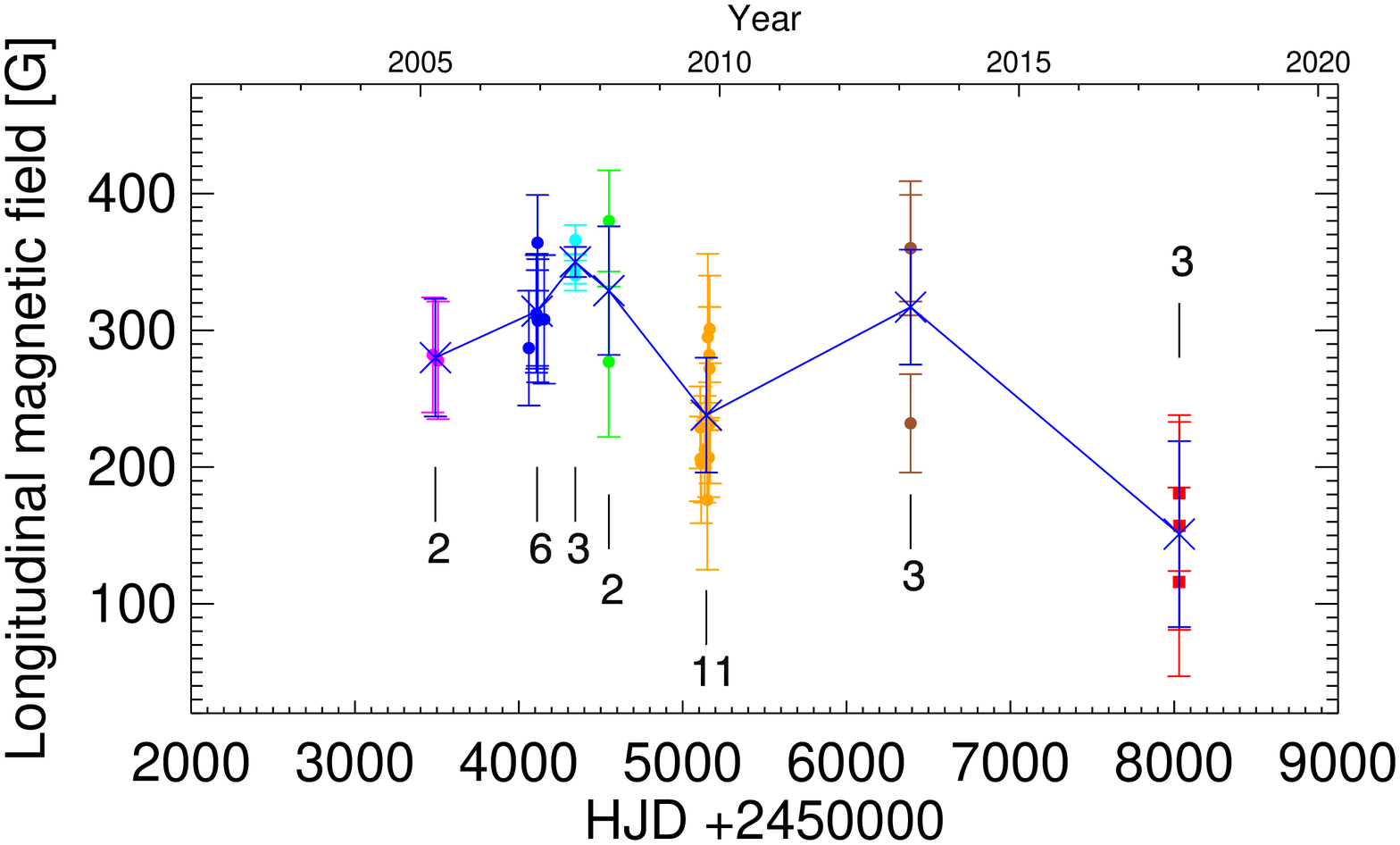}
  \caption{
    Temporal evolution of the longitudinal magnetic field. The time is
    given both in HJDs and years. Measurements belonging to different time
    bins are plotted with different colors. Identical colors are used in
    Fig.~\ref{fig:Bphased} for the phase folded points. The number of
    individual measurements in each bin is given in the plot. The blue line
    connects the time binned averages.
           }
   \label{fig:Btime}
\end{figure}

\begin{figure}
  \centering
  \includegraphics[width=.45\textwidth]{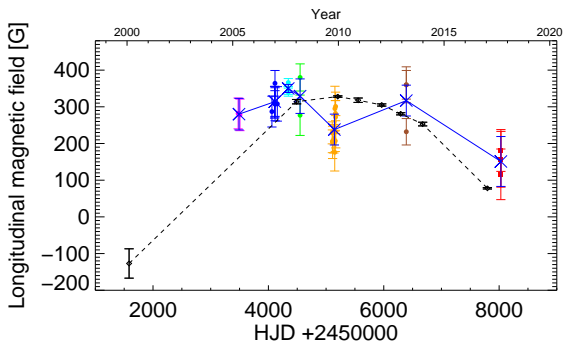}
  \caption{
    The same as Fig.~\ref{fig:Btime}. The black diamonds represent the annual
    weighted mean $\left<B_{\rm z}\right>$ measurements by
    \citet{shultz},
    connected by the dashed line.
           }
   \label{fig:Bmimes}
\end{figure}

To investigate the temporal evolution of the magnetic field strength in our 
data, we used time binned averages of the measured field strengths between 
2005 and 2017. The distribution of the binned average field strengths and the 
number of measurements used in each bin is shown in Fig.~\ref{fig:Btime} and 
the comparison of our field strength distribution with that of
\citet{shultz}
is shown in Fig.~\ref{fig:Bmimes}. In contrast to the measurements presented
by Shultz et al., we observe in both figures a slight decrease of the magnetic
field strength in 2009. Comparing the identical colors between
Figs.~\ref{fig:Bphased} and \ref{fig:Btime}, it is obvious that the observed
magnetic field decrease in the binned data is due to the rotational modulation
of the magnetic field.

%---------------------- F3

%---------------------- F4

\section{Discussion}
\label{sect:disc}

Insufficient knowledge of the strength, geometry, and time variability of
magnetic fields in hot pulsating stars prevented until now important
theoretical studies on the impact of magnetic fields on stellar rotation,
pulsation, and element diffusion.
\citet{morel2006, morel2008}
performed a NLTE abundance analysis of a sample of early-type B dwarfs with
detected magnetic fields and low $v \sin i$--values, among them $\xi^{1}$\,CMa.
Their analysis strongly supported the existence  of a population of
nitrogen-rich and boron-depleted early B-type stars. It is of interest that
the obtained abundance patterns for this group of stars are well-reproduced
by the theoretical models of
\citet{heger},
but only assuming initial velocities exceeding $\sim$200\,km\,s$^{-1}$ which
would imply strong rotational mixing and surface nitrogen enrichment. The
existence of a rotation period of the length of more than 30\,years in
$\xi^{1}$\,CMa would suggest an extreme loss of angular momentum over the
fractional main-sequence lifetime of 0.76 reported in the study of
\citet{Hubrig2006}.
Very long rotation periods of the order of decades are currently only known
for a few chemically peculiar A-type (Ap) stars with kG magnetic fields. As
evolutionary changes of the rotation periods of Ap stars during their
main-sequence lifetime are at most of the order of a factor 2
\citep{Hubrig2007},
it is believed that the period differentiation took place already before Ap
stars arrive at the main sequence. The typical rotation periods of more
massive early B-type stars with much shorter main-sequence lifetimes are
usually of the order of a few days and less and no early B-type star is known
to have long rotation periods of the order of years or more.

While rotational modulation with a short period of 2.18\,d was previously
clearly detected in FORS\,1/2 data, the rotation period longer than 30 years
reported by
\citet{shultz}
based on high-resolution spectropolarimetric observations requires additional
observational constraints. First of all, the impact of the strong pulsational
variability on the magnetic field measurements in high-resolution
spectropolarimetric observations has to be taken into account. Apart from
spectropolarimetric monitoring of $\xi^{1}$\,CMa in the coming years using
circularly polarised spectra, a useful approach to proof the presence of an
extremely long rotation could be to measure broad-band linear polarisation,
which is caused by different saturation of the $\pi$ and $\sigma$ components
of a spectral line in the presence of a magnetic field. This differential
effect is qualitatively similar for all lines, so that in broad-band
observations the contributions of all lines add up. A model for the
interpretation of such observations developed by
\citet{Landolfi1993}
does not allow to derive the field strength, but provides very useful
constraints on the geometry of the magnetic field.

\section*{Acknowledgements}
We would like to thank the anonymous referee for his useful comments.
Based on observations made with ESO Telescopes at the La Silla Paranal
Observatory under programme ID 0100.D-0110.

%% The Appendices part is started with the command \appendix;
%% appendix sections are then done as normal sections
%% \appendix

%% \section{}
%% \label{}

\section*{References}

%% If you have bibdatabase file and want bibtex to generate the
%% bibitems, please use
%%
%\bibliographystyle{elsarticle-ms} 
\bibliographystyle{aa} 
\bibliography{CMa}

%% else use the following coding to input the bibitems directly in the
%% TeX file.

%% \begin{thebibliography}{00}

%% \bibitem[Author(year)]{label}
%% Text of bibliographic item

%\bibitem[ ()]{}

%% \end{thebibliography}
\end{document}